\documentclass[twocolumn,aps,prd,groupedaddress,nofootinbib,showpacs]{revtex4-1}
\usepackage[T1]{fontenc} 
\usepackage{graphicx}
\graphicspath{{fig/}}
\usepackage{amsmath}
\usepackage{bm}
\usepackage{slashed}
\usepackage{epstopdf}
\usepackage{bbold}
\usepackage{amsfonts}

\begin{document}


\title{Quarkonium Production in an Improved Color Evaporation Model}

\author{Yan-Qing~Ma$^{1,2,3}$}
\email{yqma@pku.edu.cn}
\author{Ramona Vogt$^{4,5}$}
\email{rlvogt@lbl.gov}
\affiliation{
$^{1}$School of Physics and State Key Laboratory of Nuclear Physics and
Technology, Peking University, Beijing 100871, China\\
$^{2}$Center for High Energy Physics,
Peking University, Beijing 100871, China\\
$^{3}$Collaborative Innovation Center of Quantum Matter,
Beijing 100871, China\\
$^{4}$Nuclear and Chemical Sciences Division,
Lawrence Livermore National Laboratory,
Livermore, CA 94551, USA\\
$^{5}$Physics Department, University of California at Davis,
Davis, CA 95616, USA
}

\date{\today}

\begin{abstract}
We propose an improved version of the color evaporation model to describe
heavy quarkonium production.  In contrast to the traditional color evaporation
model, we impose the constraint that the invariant mass of the intermediate
heavy quark-antiquark pair to be larger than the mass of produced quarkonium.
We also introduce a momentum shift between heavy quark-antiquark pair and the
quarkonium. Numerical calculations show that our model can describe the
charmonium yields as well as ratio of $\psi^\prime$ over $J/\psi$ better than
the traditional color evaporation model.
\end{abstract}


\maketitle

\section{Introduction}

The study of heavy quarkonium production is one of the best ways to understand
hadronization in QCD. Currently, the most widely used theory for heavy
quarkonium production is the nonrelativistic QCD (NRQCD) approach 
\cite{Bodwin:1994jh} proposed in 1994. By introducing a systematic velocity
expansion, this theory can naturally solve the infrared divergence problem
encountered in the color singlet model (CSM) \cite{Ellis:1976fj,Carlson:1976cd,Chang:1979nn}. In this sense, NRQCD
factorization can be thought of as a generalized version of CSM. Furthermore,
it also successfully explained the $\psi'$ surplus found at Tevatron
\cite{Braaten:1994vv} by including color octet contributions.

Nevertheless, recent studies have shown that NRQCD factorization encouters
serious difficulties \cite{Brambilla:2010cs}. First, naive power counting
implies that $\psi(nS)$ and $\Upsilon(nS)$ productions at hadron colliders are
dominated by the $^3S_1$ color octet channel which results in transverse
polarization at high transverse momentum, $p_T$.  However, experimental
measurements found these states to be almost unpolarized. Current explanations
of $J/\psi$ polarization include $^1S_0$ color octet dominance
\cite{Chao:2012iv,Bodwin:2014gia,Faccioli:2014cqa} and cancelation of
transverse polarization between the $^3S_1$ and $^3P_J$ color octet channels
\cite{Chao:2012iv,Han:2014jya,Zhang:2014ybe}. Whether these explanations can
be generalized to other quarkonium states is still in question. Second, the
nonperturbative color octet long-distance matrix elements (LDMEs) extracted
from hadron colliders \cite{Ma:2010yw,Butenschoen:2010rq,Gong:2012ug} are
inconsistent with the upper bound set by $e^+ e^-$ collisions
\cite{Zhang:2009ym}.  Thus the LDMEs are not universal. Finally, there is
still no convincing proof of NRQCD factorization to all orders in $\alpha_s$.
The state-of-art proof is only to next-to-next-to-leading order for special
cases \cite{Nayak:2006fm}.

Considering the above difficulties, one should definitely study NRQCD
factorization in more detail, but, at the same time, one may need to turn to
other theories of quarkonium production. A theory which
is known to satisfy all-order factorization is
the color evaporation model (CEM) \cite{Fritzsch:1977ay,Halzen:1977rs}. In this model, to produce a charmonium
states $\psi$, one first produces a charm quark-antiquark pair $c\overline{c}$
with invariant mass smaller than the $D$-meson threshold.  The pair then
hadronizes to the $\psi$ by randomly emitting soft particles\footnote{We refer
to them as soft gluons here.}.
The production cross section is
expressed as
\begin{align}
\frac{d\sigma_\psi(P)}{d^3P} = F_\psi \int_{2m_c}^{2M_D} d M \frac{d\sigma_{c\bar c}
(M,P)}{ d M d^3 P},
\end{align}
where $m_c$ ($M_D$) is the mass of charm quark ($D$ meson) and $M$ is the
invariant mass of the $c\bar{c}$ pair. In this model, it is assumed that the
$\psi$ momentum, $P$, is approximately the same as the momentum of the
$c\overline{c}$ pair. The predictive power of the CEM is based on the
assumption that the hadronization factor $F_\psi$ is universal and thus
independent of the kinematics and spin of the $\psi$, as well as the
production process.

Although CEM is intuitive, simple, and successful to explain $J/\psi$ production data, it has very fatal flaw. A straightforward conclusion from the CEM is that the ratio of differential
cross sections of two charmonia states is independent of the kinematics
and independent of the colliding species. However, it has long known that experimental results of ratio of production cross section of $\psi^\prime$ over that of $J/\psi$ depend on their transverse momentum (recent experimental data see Refs.~\cite{Adare:2011vq,Aaij:2012ag}). This disagreement is regarded as the main evidence that CEM is a wrong.

Considered the advantages of CEM mentioned above, we may need to study whether a modification of CEM can provide a correct theory for quarkonium production. In this paper, by taking into account physical effects overlooked in the
original CEM, we propose an improved color-evaporation model (ICEM).  On the one hand,
the nice features of CEM are retained in the ICEM, including having only one
parameter for each quarkonium state 
and satisfying all-order factorization. On the other hand, the ICEM can correctly describe charmonium production cross section ratios.

\section{The improved color-evaporation model}

\begin{figure}
\begin{center}
\includegraphics[width=0.4\textwidth]{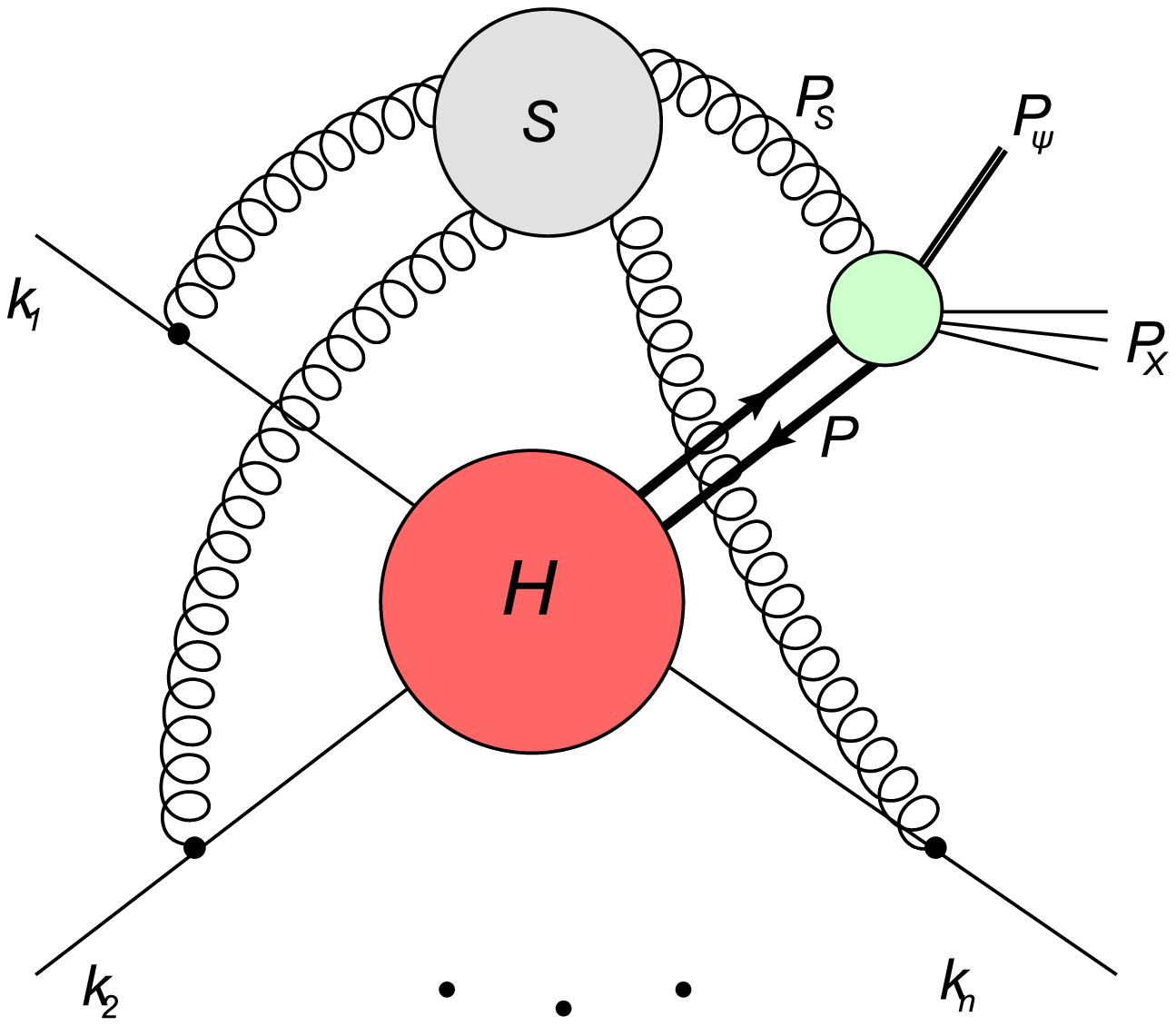}
\end{center}
\caption{An illustration of charmonium production in a high energy collision.
See text for details.}
\label{fig:picture}
\end{figure}

Our picture of heavy quarkonium (say charmonium) production is as follows.
To produce a charmonium state $\psi$, it is necessary to produce a
$c\overline{c}$ pair in the hard collision, because the mass of
the $c\overline{c}$ pair is much larger than the QCD nonperturbative scale
$\Lambda_{\text{QCD}}$. Before the $c\overline{c}$ pair hadronizes to charmonium,
it will exchange many soft gluons between various color sources, as well as
emit soft gluons.  An illustration of this picture is given in
Fig.~\ref{fig:picture}. In this figure, the blob marked by `$H$' denotes the
hard collision kernel, the blob marked by `$S$' denotes soft interactions, and
the thick double lines denotes the $c\overline{c}$ pair with momentum $P$. To
separate the hard part from the other parts, we introduce a scale $\lambda$
with $m_c \gg \lambda \gg \Lambda_{\text{QCD}}$, and define the hard part as all
particles that are off shell by more than $\lambda^2$.

We emphasize that we distinguish soft gluons exchanged between the
$c\overline{c}$ pair and other color sources (with momentum denoted by $P_S$)
from soft gluons emitted by the $c\overline{c}$ pair (with momentum denoted by
$P_X$).  Indeed, these two kinds of soft gluons are significantly different.
The total
energy of exchanged gluons can be either positive or negative.  However, the
emitted gluons will eventually evolve to experimentally observable particles.
Thus their total momentum must be time-like and their total energy must be
positive.

In our model, we construct a relationship between $P$ and
$\langle P_\psi \rangle$, the average momentum of $\psi$ that has hadronized
from a $c\overline{c}$ pair with fixed momentum $P$. The relationship is easy
to obtain in the rest frame of $P$, with $P=(M, 0, 0, 0)$. For each event, we
have
\begin{align}
P=P_\psi + P_S + P_X \, \, .
\end{align}
In the spirit of the traditional CEM, we assume the distributions of $P_S$ and
$P_X$ are rotation invariant in this frame, which implies $\langle P_S \rangle=(m_S,0,0,0)$
and $\langle P_X \rangle=(m_X,0,0,0)$. Because exchanged gluons can flow in
either direction, we may expect $m_S\approx 0$.  Thus
$\langle P_\psi \rangle=(M-m_X,0,0,0)$ with $m_X>0$. Therefore,
\begin{align}\label{eq:Mlimit}
M_\psi<M-m_X<M \, \, ,
\end{align}
where we use the fact that $\langle P_\psi^0 \rangle$ must be larger than
$M_\psi$. Equation~\eqref{eq:Mlimit} sets a lower limit on $M$ that is
significantly different from the lower limit $2m_c$ of the traditional CEM.

As both $P_S$ and $P_X$ are order of $\lambda$, power counting of $P_\psi$ gives
$(O(m_c),O(\lambda),O(\lambda),O(\lambda))$. Combining with the on-shell
condition $P_\psi^2=M_\psi^2$, we arrive at $P_\psi^0=M_\psi+O(\lambda^2/m_c)$.
Thus we have
\begin{align}\label{eq:Plimit}
\langle P_\psi \rangle=\frac{M_\psi}{M}P+O(\lambda^2/m_c) \, \, ,
\end{align}
which again differs from the relation used in the traditional CEM where
$P_\psi$ is identified with $P$. Note that the proportionality between the
momenta of the mother and daughter particles in Eq.~\eqref{eq:Plimit} was
first proposed in Ref.~\cite{Ma:2010vd} to relate the momentum of the
$\chi_{cJ}$ and the $J/\psi$ produced by its decay.  It has since been used in
many calculations of quarkonium production in the NRQCD framework. In this
paper, we prove the relation rigorously with clear assumptions. By combining
Eqs.~\eqref{eq:Mlimit} and \eqref{eq:Plimit}, we arrive at the improved
color evaporation model (ICEM):
\begin{align}\label{eq:model}
\begin{split}
\frac{d\sigma_\psi(P)}{d^3P} &= F_\psi \int_{M_\psi}^{2M_D} d^3P' d M \frac{d\sigma_{c\bar c}(M,P')}{ d M d^3 P'} \delta^3(P-\frac{M_\psi}{M} P')\\
&= F_\psi \int_{M_\psi}^{2M_D} d M \frac{d\sigma_{c\bar c}(M,P'=(M/M_\psi) P)}{ d M d^3 P},
\end{split}
\end{align}
with correction at $O(\lambda^2/m_c^2)$.
If one is only interested in the transverse momentum distribution, we have
\begin{align}
\begin{split}
\frac{d\sigma_\psi(P)}{dp_T} = F_\psi \int_{M_\psi}^{2M_D} d M \frac{M}{M_\psi}
\frac{d\sigma_{c\bar c}(M,P')}{ d M d p'_T}|_{p'_T=(M/M_\psi) p_T}.
\end{split}
\end{align}

Before performing any numerical calculations, we can already expect some
advantages of the ICEM. First, because there is an explicit charmonium mass
dependence in Eq.~\eqref{eq:model}, the ratio of differential cross sections of
two charmonia is no longer $p_T$-independent in the ICEM. Thus it is possible
to explain data such as $d\sigma_{\psi(2S)}/d\sigma_{J/\psi}$. Second, by
making a distinction between the momentum of the $c\overline{c}$ pair and that
of charmonium, the predicted $p_T$ spectra will be softer and thus may
explain the high $p_T$ data better.

We emphasize that the ICEM Eq.~\ref{eq:model} does not mean that $c\bar c$ pair with invariant mass smaller than $M_\psi$ has no possibility to hadronize to $\psi$. In fact, this kind of $c\bar c$ pair can absorb energy by interacting with other color source, and thus can have larger invariant mass and hadronize to $\psi$. At the same time, even if the invariant mass of $c\bar c$ pair is larger than $M_\psi$, it may loss energy by interacting with other color source, and eventually cannot hadronize to $\psi$ because of invariant mass being too small. By assuming $m_S\approx0$, we effectively approximate that the two effects cancel each other.  As a result, the Eq.~\ref{eq:model} should be only interpreted at the integration level.

An exception for the above argument is for the ground state particle production, say $\eta_c$ for charmonium. Based on the quark-hadron duality, $c\bar c$ pair with invariant mass smaller than $D$ meson threshold must hadronize to charmonium, therefore it is not possible for a $c\bar c$ pair with $M_{c\bar c} > M_{\eta_c}$ to emit too much energy so that its invariant mass becomes smaller than $M_{\eta_c}$. This means that the approximation $m_S\approx0$ is not reasonable here and thus ICEM is not good for $\eta_c$ production. However, for $\eta_c$ production, as the condition Eq.~\eqref{eq:Mlimit} is not needed, the original CEM should be good.

\section{Numerical results}

\begin{figure*}
\begin{center}
\includegraphics[width=0.4\textwidth]{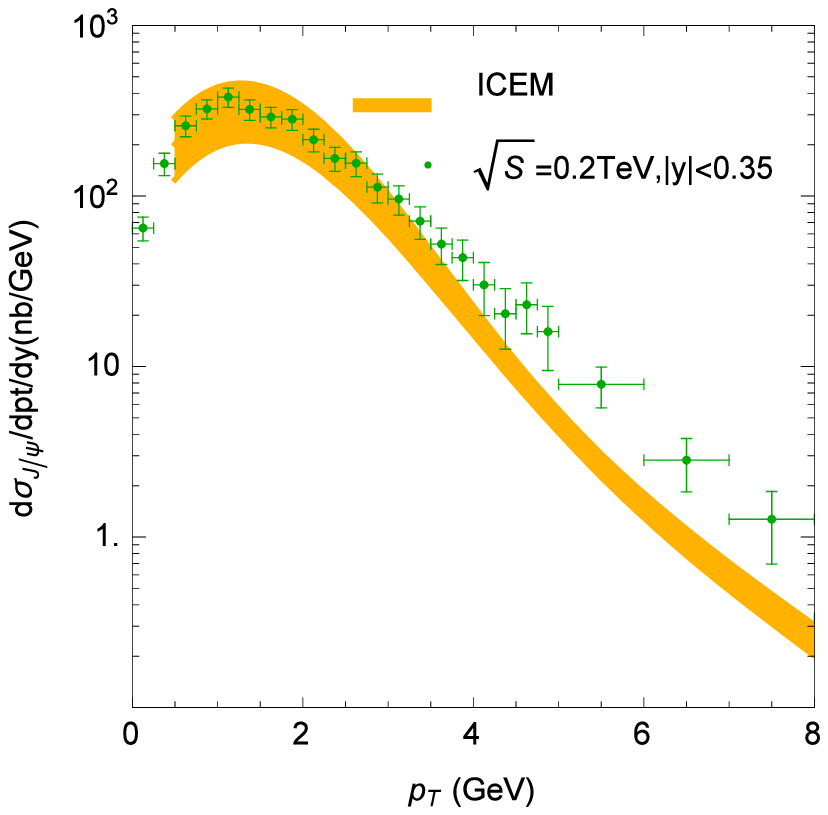}
\includegraphics[width=0.4\textwidth]{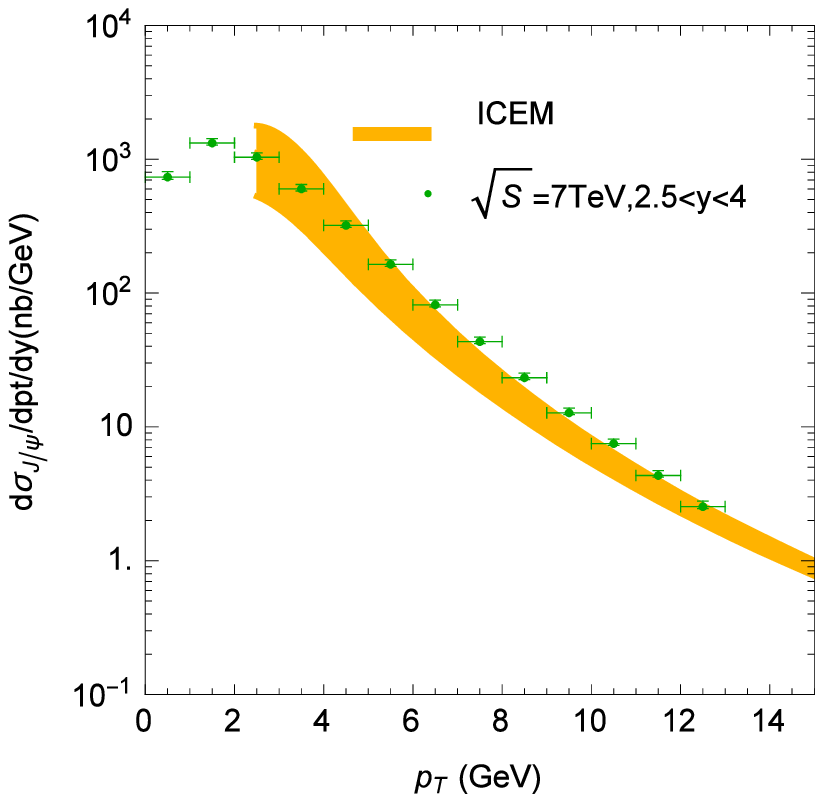}
\end{center}
\caption{Results for $J/\psi$ production. The 0.2 TeV PHENIX data
and 7 TeV LHCb data are taken from Ref.~\cite{Adare:2011vq} and
Ref.~\cite{Aaij:2011jh}, respectively.}
\label{fig:jpsi}
\end{figure*}

\begin{figure*}
\begin{center}
\includegraphics[width=0.4\textwidth]{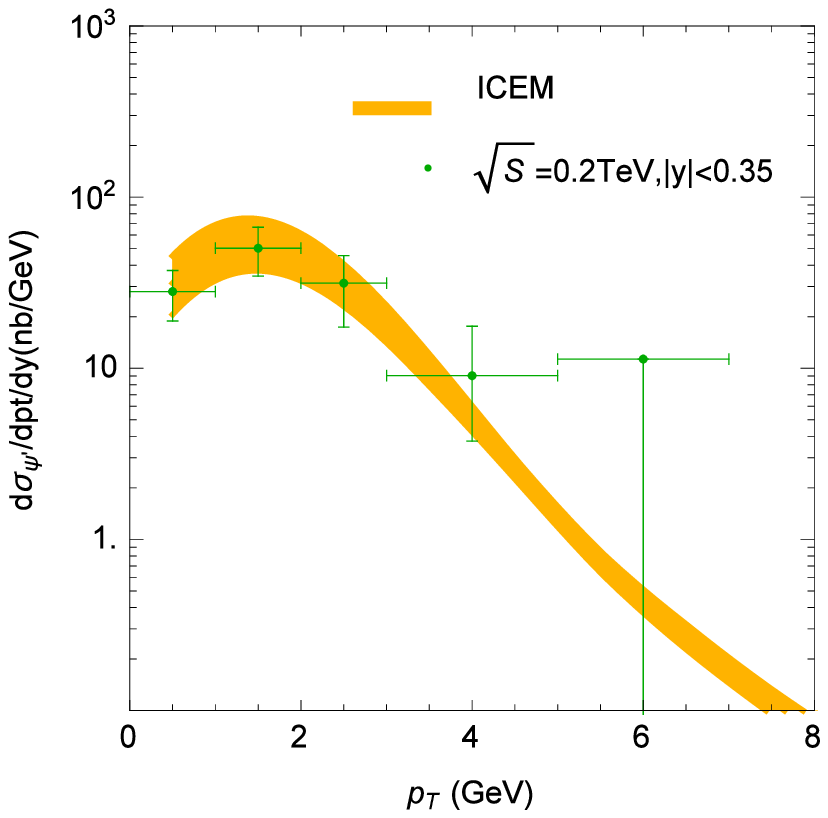}
\includegraphics[width=0.4\textwidth]{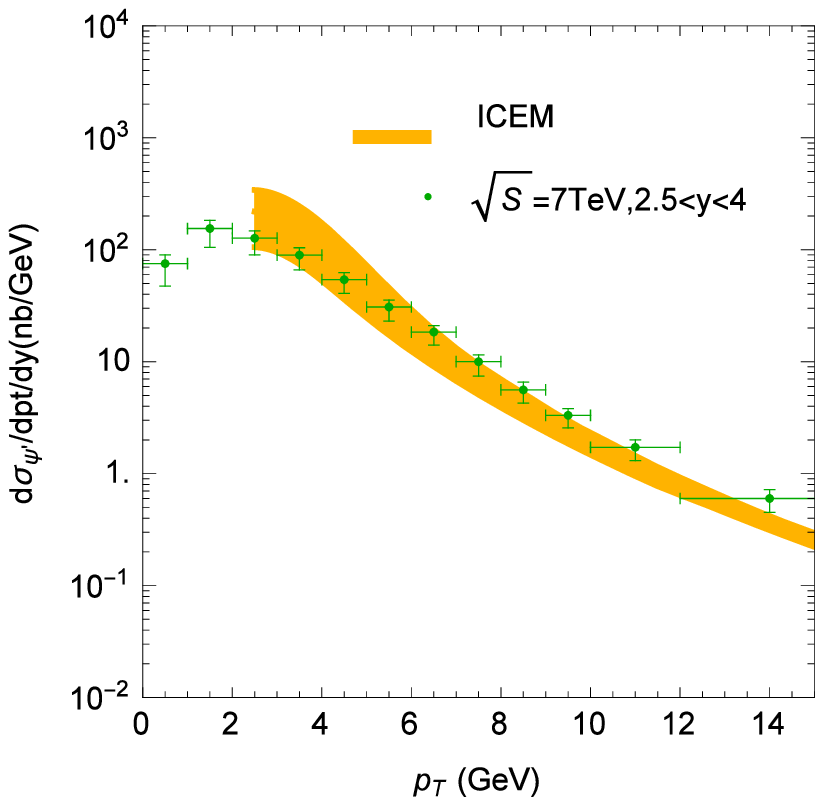}
\end{center}
\caption{Results for $\psi^\prime$ production. The 0.2 TeV PHENIX data and 7 TeV
LHCb data are taken from Ref.~\cite{Adare:2011vq} and Ref.~\cite{Aaij:2012ag},
respectively.}
\label{fig:psi2s}
\end{figure*}

\begin{figure*}
\begin{center}
\includegraphics[width=0.4\textwidth]{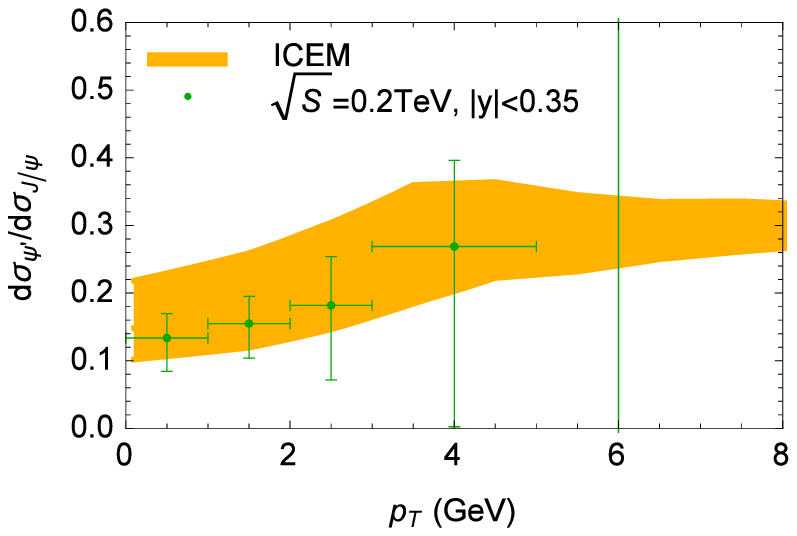}
\includegraphics[width=0.4\textwidth]{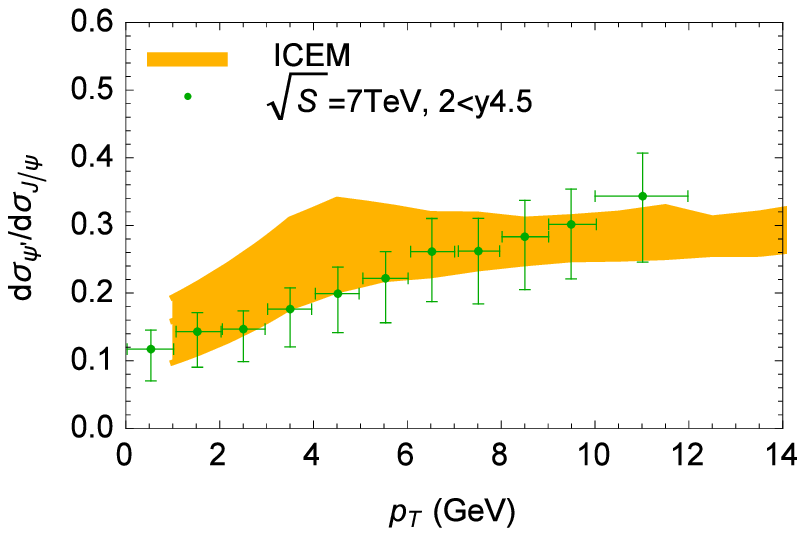}
\end{center}
\caption{Results for ratio of the $\psi^\prime$ production cross section to
that of $J/\psi$. The 0.2 TeV PHENIX data and 7 TeV LHCb data are taken from
Ref.~\cite{Adare:2011vq} and Ref.~\cite{Aaij:2012ag}, respectively.}
\label{fig:ratio}
\end{figure*}

To confront our model with experimental data, we updated the CEM parameters
determined in Ref.~\cite{Nelson:2012bc}.  In that work, in an attempt to reduce
the uncertainty on the total charm cross section, the charm mass was fixed at
$1.27 \pm 0.09$~GeV while the factorization and renormalization scales were fit
to a subset of the measured total charm cross section data.  The values found
were $\mu_F/m = 2.1^{+2.55}_{-0.85}$ and $\mu_R/m = 1.6^{+0.11}_{-0.12}$ employing
the CT10 proton parton densities \cite{Lai:2010vv}.

The central open charm parameter set $(m,\mu_F/m,\mu_R/m) = (1.27, 2.1, 1.6)$
was used to calculate the energy dependence of the forward $J/\psi$ cross
section, $\sigma(x_F > 0)$, in the CEM using the exclusive $c \overline c$
production code described in Ref.~\cite{Mangano:1991jk}.  Because the NLO
$c \overline c$ code is an exclusive calculation, the mass cut is on the
invariant average over kinematic variables of the $c$ and $\overline c$. Thus,
in this calculation $\mu_F$ and $\mu_R$ are defined relative to the transverse
mass of the charm quark, $\mu_{F,R} \propto m_T = \sqrt{m^2 + p_T^2}$ where
$p_T^2 = 0.5(p_{T_c}^2 + p_{T_{\overline c}}^2)$.
The normalization $F_\psi$ is
the scale factor that adjusted the fraction of the total charm cross section in
the mass range $2m < M < 2m_D$ to the forward cross section data.

To determine
the uncertainty on the $J/\psi$ calculation, the charm mass was varied between
the upper and lower limits, 1.36 and 1.18 GeV respectively, for the central
values of $\mu_F/m$ and $\mu_R/m$, and the scales were varied around their
central values while the charm mass was held fixed at its central value of
1.27 GeV: $(\mu_F/m,\mu_R/m) = (C,L), (L,C), (L,L), (C, H), (H, C), (H,H)$
where $H (L)$ is the upper (lower) limit of the factorization and
renormalization scales determined from the charm fits.  Using the same value
of $F_\psi$ in all cases, the uncertainty band on the $J/\psi$ cross section
was calculated by finding the upper and lower limits of the mass and scale
variations and adding them in quadrature, as discussed in
Refs.~\cite{Nelson:2012bc,Cacciari:2005rk}.

To calculate the charmonium $p_T$ dependence, a Gaussian transverse momentum
broadening is added to the final state.  The value of the average $k_T$ kick
applied was taken to be
$\langle k_T^2 \rangle = 1 + (1/12)\ln(\sqrt{s}/20)$~GeV$^2$
\cite{Nelson:2012bc}, giving 1.19~GeV$^2$ at RHIC and 1.49~GeV$^2$ at 7 TeV.

Since the ICEM calculation discussed here reduces the cross section relative
to the calculation in Ref.~\cite{Nelson:2012bc}, the value of $F_\psi$ had to
be increased by 40\% to retain agreement with the data.  The $\psi'$ cross
section and its uncertainty was calculated with the same parameters but with
a value of $F_{\psi^\prime}$ scaled to the $\psi'$ data.

To obtain the uncertainty on the $\psi'/\psi$ ratio, the mass and scale
uncertainties were assumed to be correlated.  The resulting uncertainty band is
dominated by the scale uncertainty, the mass uncertainty is small.

Our results for $J/\psi$ production cross section as a function of $p_T$ are
shown in Fig.~\ref{fig:jpsi}, where we compare with data at hadron colliders
for center of mass energies of 0.2 TeV and 7 TeV. The 0.2 TeV RHIC data are
measured by the PHENIX Collaboration \cite{Adare:2011vq} at central rapidities,
$|y| < 0.35$, and the 7 TeV LHC data are measured by the LHCb Collaboration
\cite{Aaij:2011jh} at forward
rapidity, $2.5 < y < 4$.  The largest discrepancy between the model and the
data is in the RHIC data at intermediate $p_T$, $4 < p_T < 7$ GeV. However,
since the experimental uncertainty is rather large in this region, our results
are in general agreement with the data.

We now turn to the $\psi^\prime$ production cross section as a function of $p_T$
in Fig.~\ref{fig:psi2s}.   We again compare with the midrapidity PHENIX data
\cite{Adare:2011vq} at 0.2 TeV and the forward LHCb data \cite{Aaij:2012ag} at
7 TeV.  Since the $\psi^\prime$ rates are generally lower, the measured
uncertainty is larger.  Given this, the agreement of the calculation with the
data is also good.

The ratio of the production cross sections of $\psi^\prime$ to that of $J/\psi$
as a function of $p_T$ is given in Fig.~\ref{fig:ratio}. The 0.2 TeV RHIC data
and 7 TeV LHC data are taken from Ref.~\cite{Adare:2011vq} and
Ref.~\cite{Aaij:2012ag}, respectively.  Although the original CEM predicts a
constant for this ratio, in contradiction with the data, our ICEM calculations
are in good agreement with all data.

\section{Summary and discussion}

By distinguishing between exchanged and emitted soft gluons and considering
some physical constraints, we propose an improved color evaporation model for
charmonium production. Comparison with data shows that the ICEM can nicely
reproduce the $p_T$ dependence of the ratio of the $\psi^\prime$ to $J/\psi$
production cross sections. Thus, this improved model overcomes one of the main
obstacles of the original CEM. The success of the ICEM calculation confirms
our picture of charmonium production.

We note that the question of polarization in the ICEM as well as the original
CEM has not yet been addressed. As seen in the NRQCD approach, the polarization
is an important test of models.  The prediction of the final-state charmonium
polarization depends on whether soft gluons change spin and angular momentum
of the $c\bar c$ pair. A preliminary study of charmonium polarization in the CEM will be presented
elsewhere \cite{CheungVogt}.

\begin{acknowledgments}

We thank Kuang-Ta Chao, Raju Venugopalan and Hong-Fei Zhang for useful discussions.
The work of RV was performed under the auspices of the
U.S. Department of Energy by Lawrence Livermore National Laboratory under
Contract DE-AC52-07NA27344and and supported by the U.S. Department of Energy,
Office of Science, Office of Nuclear Physics (Nuclear Theory) under contract
number DE-SC-0004014.
.

\end{acknowledgments}

\providecommand{\href}[2]{#2}\begingroup\raggedright\endgroup


\end{document}